\documentclass{emulateapj}

\newcommand{\etal}{{\it et al.}}

\begin{document}

\title{Photometric Redshifts and Photometry Errors}

\shorttitle{Photometric Redshifts and Photometry Errors}

\author{D. Wittman, P. Riechers, and V.~E. Margoniner\altaffilmark{1}}
\affil{Physics Department, University of California, Davis,
  CA 95616; dwittman@physics.ucdavis.edu}
\altaffiltext{1}{Current address: Physics Department, California State
University, Sacramento, CA 95819}

\keywords{surveys---galaxies: photometry---methods: statistical}

\begin{abstract}
We examine the impact of non-Gaussian photometry errors on photometric
redshift performance.  We find that they greatly increase the scatter,
but this can be mitigated to some extent by incorporating the correct
noise model into the photometric redshift estimation process.
However, the remaining scatter is still equivalent to that of a much
shallower survey with Gaussian photometry errors.  We also estimate
the impact of non-Gaussian errors on the spectroscopic sample size
required to verify the photometric redshift rms scatter to a given
precision.  Even with Gaussian {\it photometry} errors, photometric
redshift errors are sufficiently non-Gaussian to require an order of
magnitude larger sample than simple Gaussian statistics would
indicate.  The requirements increase from this baseline if
non-Gaussian photometry errors are included.  Again the impact can be
mitigated by incorporating the correct noise model, but only to the
equivalent of a survey with much larger Gaussian photometry errors.
However, these requirements may well be overestimates because they are
based on a need to know the rms, which is particularly sensitive to
tails.  Other parametrizations of the distribution may
require smaller samples.
\end{abstract}

\section{Introduction}

Photometric redshifts (Connolly \etal\ 1995, Hogg \etal\ 1998, Benitez
2000) are of increasing importance in observational tests of
cosmology.  Predicting photometric redshift performance has therefore
become an important part of planning large optical surveys.  There are
two distinct aspects of performance to consider.  First, there are
straightforward goals of accuracy and precision.  Second, to control
systematic errors in the downstream science, one must be able to {\it
  know}, in some cases rather stringently, the accuracy and precision
of the photometric redshifts in the actual survey (Ma \etal\ 2006,
Huterer \etal\ 2006).  Knowing the actual photometric redshift
precision can be more important than maximizing the precision.  For
example, cosmic shear tomography calls for relatively wide redshift
bins ($dz \sim 0.2$).  Leakage between bins, to the extent that it is
known, can be precisely incorporated into comparisons between models
and data.  This by itself is not very demanding in terms of
photometric redshift precision.  However, in a large survey with very
small statistical errors, the leakage must be known very precisely to
avoid nontrivial systematic errors.  Ma \etal\ (2006) estimate that
for cosmic shear tomography with next-generation surveys, the bias and
rms scatter in each redshift bin must be known to $\sim$0.003 to avoid
degrading the shot-noise-limited constraints on dark energy.

To first order, photometric redshift performance depends on filter
set, signal-to-noise (S/N), and the desired range of redshifts and
galaxy types.  Here we wish to call attention to an often overlooked
aspect: photometry errors.  Photometric redshift simulations and
real-life implementations typically assume Gaussian photometry errors.
Real data are more complicated. As one anecdote, Cameron \& Driver
(2007) note that in one catalog of 42 galaxies with both photometric
and spectroscopic redshifts, there were six outliers, all of which
had questionable photometry due to saturation, neighbors, or multiple
nuclei.  In this paper we show that knowing the true distribution of
errors is important for optimizing photometric redshift precision. We
also discuss how that in turn affects the size of the spectroscopic
sample required to characterize the photometric redshift errors in a
survey.

\section{Methods}

We conduct four sets of simulations built around the following basic
setup.  We use the Bayesian Photometric Redshift (BPZ, Benitez 2000)
code, which uses a set of template galaxy spectral energy
distributions (SEDs) and a set of priors to help break degeneracies in
color space.  We chose the six SED templates and the HDFN prior
detailed in Benitez (2000).  BPZ is representative of one of two types
of methods in the photometric redshift community.  We discuss possible
impacts on the other type, training-set methods, in
\S\ref{sec-discussion}.  The choice of filter set is not important for
this demonstration.  We use the same filter set (F300W, F450W, F606W,
F814W, J, H, K) used for the Hubble Deep Field North (HDFN)
photometric redshifts discussed in Hogg \etal\ (1998), Benitez (2000),
and Fernandez-Soto \etal\ (1999, 2001).

Each simulation generates a synthetic catalog of 6000 galaxies evenly
spread throughout the F814W magnitude range 20--26.  This and other
aspects of the simulations are not realistic, but are adopted to
facilitate analysis by covering parameter space evenly.  The results
presented here therefore do not apply quantitatively to any real
survey, but they demonstrate the issues.  The simulator uses each
galaxy's magnitude to choose a random type and redshift following the
distributions described by the priors.  It then looks up the synthetic
observer-frame colors of that type at that redshift, and adds noise
(the character of which varies with the simulation) before saving the
catalog.  An unrealistic aspect of the noise in all simulations is
that it is a fixed percentage of the model flux.  That is, every
galaxy is observed at the same S/N, regardless of magnitude, redshift,
or filter.  This is another analysis convenience.  The effect of
varying S/N was explored in one specific case by Margoniner \& Wittman
(2007), and will have to be customized to each survey.

We then run the catalogs through BPZ, with the HDFN prior turned on,
and analyze the performance in terms of 
$\delta z \equiv {z_{\rm phot} - z_{\rm spec} \over 1+ z_{\rm spec}}$, 
specifically the bias
$\bar{\delta z}$ and the scatter $\delta z_{rms}$.

\section{Realizations}

\begin{figure}
\centerline{\resizebox{3in}{!}{\includegraphics{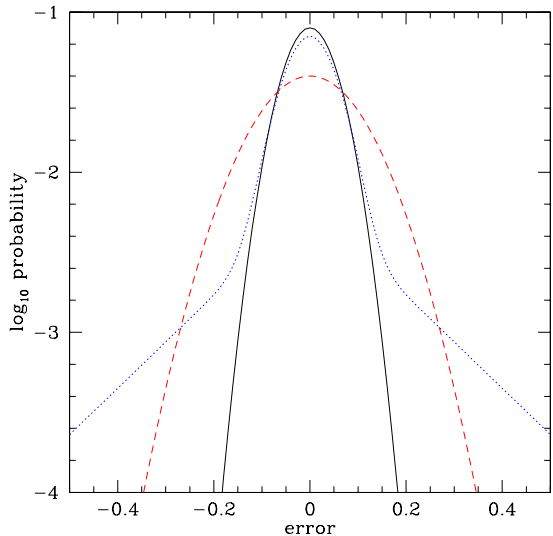}}}
\caption{Photometry error distributions, SIM1: 5\% Gaussian (solid
  black curve); SIM2: 10\% Gaussian (dotted red curve); SIM3 and SIM4:
  5\% Gaussian with exponential tails (dashed blue curve).
\label{fig-phot}}
\end{figure}

\begin{figure}
\centerline{\resizebox{3in}{!}{\includegraphics{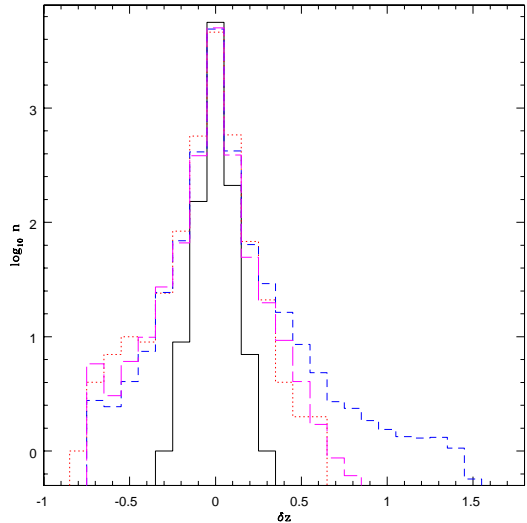}}}
\caption{Distributions of $\delta z$: colors and linetypes are as in
  previous figure, with the addition of SIM4 (long-dash magenta
  curve), which uses the non-Gaussian noise model in the photometric
  redshift estimation.
\label{fig-dz}}
\end{figure}

As baselines, we do two simulations with Gaussian noise: SIM1 with 5\%
noise ($S/N=20$) and SIM2 with 10\% noise ($S/N=10$).  These
photometry error distributions are shown in Figure~\ref{fig-phot}.
The resulting $\delta z$ distributions are shown in
Figure~\ref{fig-dz}.  In both cases, the bias is small (0.003 or less
in absolute value) and not inconsistent with zero.  The scatter
depends strongly on S/N: $\delta z_{rms} = 0.026$ for S/N of 20,
increasing to 0.070 for S/N of 10.  We also did a run with $S/N=100$,
not shown in the figures: $\delta z_{rms} = 0.004$.  This is extremely
tight because the quoted S/N is achieved in {\it each} band for {\it
each} galaxy.

Next, we add non-Gaussian tails to the photometry error distribution.
We adopt a functional form
$$ p(\delta f) = {1\over\sigma\sqrt{2\pi}+AB} (\exp(-{(\delta f)^2
  \over 2 \sigma^2 }) + A \exp(-{|\delta f| \over B}))$$ where $\delta
  f$ is the flux error, $\sigma$ describes the width of the Gaussian
  core, and the parameters A and B describe the tails.  For a given
  $\sigma$, the fraction of galaxies in the tails is sensitive to
  changes in the product $AB$ but relatively insensitive to changes in
  A and B as long as the product is held constant.  There is little
  published data on realistic values of A and B.  Margoniner \&
  Wittman (2007) briefly descibe photometry simulations in which
  synthetic galaxies are added to real images from the Deep Lens
  Survey (DLS, Wittman \etal\ 2002).  We roughly match the fraction of
  objects in that tail, but with two symmetric tails and $\sigma=0.05$
  as in SIM1, by setting $A=0.1$ and $B=0.15$ or $3\sigma$.  For this
  choice of A and B, used in SIM3 and SIM4 and shown as the blue dash
  curve in Fig.~\ref{fig-phot}, the tails begin to dominate over the
  Gaussian core at 2.51 times the rms of the Gaussian core, and 9.4\%
  of the galaxies are ``in'' the tails, compared to 1.2\% falling
  outside 2.51$\sigma$ for a pure Gaussian.  The rms of the
  distribution is 0.103, very close to that of SIM2.

As a comparison, the photometry error distribution for bright,
unresolved objects in the Sloan Digital Sky Survey (SDSS) is published
in Fig. 3 of Ivezi{\'c} \etal\ (2003), who state that 0.9\% of objects
lie outside of $\pm 3\sigma$ (where $\sigma=0.02$), vs. 0.3\% for a
pure Gaussian.  This observation, and the figure, are reasonably
approximated by $A=0.1$ and $B=0.0235$ or 1.2$\sigma$.  These tails
are much smaller than used in SIM3 and SIM4, which have 7.3\% of their
galaxies outside $\pm 3\sigma$.  However, the available SDSS data are
for {\it bright ($g<20.5$) point sources}.  Photometry is notably more
difficult for extended sources and for faint sources.  In the DLS
simulations, $A$ is consistent with zero for bright ($20<R<22$)
galaxies, and grows steadily with magnitude.  Of course, most of the
galaxies in a deep survey are at the faint end.  Therefore, while
noting the near-Gaussianity of the SDSS bright point-source
photometry, we believe that heavier tails are currently more
appropriate for faint galaxies in deep ground-based surveys.

We attribute the Gaussian cores of these distributions to photon
statistics, which is the nominal error reported by most photometry
packages, and the tails to other effects such as crowding.  This is a
reasonable approximation for ground-based data, with many sky photons
per pixel and galaxies usually much fainter than sky.  For space-based
photometry, crowding is less important, but photon statistics are less
Gaussian due to the smaller number of photons.  The tails in this
paper are meant to emulate ground-based surveys as described above.
We quantify their impact by estimating redshifts in SIM3 using the
nominal Gaussian photometry error as input to BPZ.  Averaged over 100
realizations, $\bar{\delta z}$ remained small (0.0038), but $\delta
z_{rms}$ increased to 0.092.  The distribution is shown in as the blue
short-dash histogram in Fig.~\ref{fig-dz}.

Clearly, these tails are very harmful.  Adding them to the $S/N=20$
distribution more than doubled $\delta z_{rms}$.  In fact, {\it
doubling} the Gaussian photometry noise had less impact on $\delta
z_{rms}$ than did adding these tails.  Surveys will have to control
the tails of their photometry error distributions if they are to reach
the photometric redshift performance expected based on their filter
set and S/N.  Modern surveys do recognize this and work to reduce the
tails, but tails will always be present at some level.  Legacy surveys
may have non-Gaussian errors frozen into their data, and new surveys
will find it expensive to eliminate all non-Gaussian sources of error.
Therefore, we investigate the extent to which knowledge of these
errors can render them less damaging to photometric redshifts.

\section{Living with Non-Gaussian Errors}

Accounting for these errors is straightforward.  In the BPZ code, the
probability of observing colors $C$ given a model SED type $T$ and
redshift $z$, $p(C|T,z)$ is simply a Gaussian of width set by the
nominal photometry errors for that galaxy.  In SIM4, we use the same
input photometry as SIM3 but replace that noise model with the full
heavy-tailed distribution used the generate the catalog.  The
resulting $\delta z$ distribution is shown in Fig.~\ref{fig-dz} as the
long-dash magenta histogram.  The outliers in $\delta z$ which
appeared in SIM3 have now largely disappeared, and $\delta z_{rms}$ is
down to 0.072.  This is comparable to $\delta z_{rms}$ in SIM2, which
had twice the simulated sky noise, but no tails.

The scatter in $\delta z$ increases to 0.082 if one uses the
unmodified BPZ code assuming Gaussian errors, but with an rms of 0.1
instead of 0.05, to roughly approximate the wider distribution of
photometry errors. As another comparison case for incorrect noise
models, we estimated redshifts from a SIM2 realization using the SIM1
noise model.  In this case, $\delta z_{rms}$ changed by only 0.003,
which was not quite significant given the sample size.  Thus, it
appears that if the photometry errors are Gaussian, knowing the width
of that Gaussian is not very critical.  We see from Fig.~\ref{fig-dz}
that it is the 1 in $\sim$500 outlier that is responsible for the poor
performance of SIM3.  SIM2 lacks extreme outliers, so qualitatively,
its better performance makes sense despite its broader core.  Yet this
degree of insensitivity to the Gaussian width is somewhat surprising.

For comparison, we perform a version of SIM4 in which the tails are
much less prominent, as in the SDSS bright point-source photometry:
$\sigma=0.05$, $A=0.1$, and $B=0.06$ (1.2$\sigma$).  We find that
$\delta z_{rms}=0.031$, with the noise model affecting only the fourth
decimal place.  The photometry tails are apparently small enough that
including them in the noise model is not very helpful, but overall
performance is still significantly worse than with no tails at
all. (SIM1 had $\delta z_{rms}=0.026$, while the variation from
realization to realization is $\sim0.001$ and these numbers are quoted
after averaging over 100 realizations.)  This indicates that even
small photometry tails can have a significant impact on photometric
redshift performance.

\section{Discussion}
\label{sec-discussion}

It is not surprising that tails in the photometry error distribution
can cause outliers in the $\delta z$ distribution.  However, a number
of points are worth remarking:
\begin{itemize}

\item Adding heavy tails (comprising $\leq 10\%$ of the galaxies)
caused more increase in $\delta z_{rms}$ than did {\it doubling} the
Gaussian photometry error.  In other words, the photometric redshift
performance of a survey with large tails could be worse than that of a
survey with {\it half} the S/N but with no tails.  Surveys should
therefore pay close attention to reducing the tails of the color
errors.  This is not the same as reducing the tails of the flux
errors.  As an extreme example, if an equal fraction of light is lost
in all filters, the colors are unaffected.

\item Assuming that non-Gaussian errors can never be entirely
  eliminated, the effect of the tails on photometric redshift
  performance can be mitigated by including an accurate noise model in
  the photometric redshift process.  This will in
  turn require extensive Monte Carlo simulations which include all
  important sources of non-Gaussian errors, such as crowding and
  complex galaxy morphology.  In addition, the importance of the tails
  is likely to vary with magnitude, seeing, etc.

\item No clear rule is evident for required accuracy of the noise
model.  Photometric redshift precision was not significantly affected
when errors and model were both Gaussian but the rms was wrong by a
factor of two.  When errors were heavy-tailed, approximating them with a
Gaussian of the same rms won back about half of the precision that
could be won back with the fully correct noise model.

\item Even very small tails have a measurable impact on $\delta
z_{rms}$, but in this case the noise model made no measurable
difference.

\end{itemize}

The tails also have a disproportionate impact on the problem of
knowing $\delta z_{rms}$ precisely for each redshift bin, whereas
precision on $\bar{\delta z}$ did not suffer substantially.  If the
$\delta z$ distribution is Gaussian, the spectroscopic sample size
required to calibrate $\delta z_{rms}$ to a desired accuracy
$\sigma_{cal}$ is $\sim {(\delta z_{rms})^2 \over 2 \sigma^2}$ (this
of course assumes that the spectroscopic sample is representative of
the photometric sample).  For $\sigma_{cal}=0.003$ and a class of
sources with $\delta z_{rms} = 0.026$ as in SIM1, only $\sim 40$
galaxies would be required.  However, bootstrap resampling of SIM1
shows that seven times more galaxies are required to know $\delta
z_{rms}$ to the same accuracy, due to its non-Gaussian tails (which
stem from the properties of galaxies in color space, not from the
photometry).  For SIM2, the factor is thirteen, presumably because the
greater noise in SIM2, although still Gaussian, allows more
near-degeneracies in color space to come into play.  For SIM3 with its
heavy photometry tails, the factor is $\sim$50.  However, this can be
much reduced simply by incorporating the correct noise model into the
photometric redshift estimation.  SIM4 requires ``only'' $\sim 25$
times as many galaxies as the Gaussian prediction would suggest, and
the Gaussian prediction is itself $\sim 2$ times smaller than for
SIM4, because of the smaller $\delta z_{rms}$.  Of course, it would be
preferable to reduce non-Gaussian tails in the underlying photometry
as much as possible, as dramatically illustrated by the large
remaining differences between SIM4 and either SIM1 or the simulation
with SDSS-like tails.

We caution that this procedure may substantially overestimate
spectroscopic sample requirements.  They are based on the Gaussian
model of photometric redshift errors employed by Ma \etal\ (2006), who
derived a prescription for precision of our knowledge of $\delta
z_{rms}$.  But the rms of a distribution is driven by its tails, so
that the tails seem to be all-important here.  If the photometric
redshift error model used in the cosmological parameter estimation
were modeled differently, the tails could assume a more proportional
influence, and fewer spectroscopic redshifts would be required to
characterize their effect.  Mandelbaum \etal\ (2007) discuss some
related aspects in the context of galaxy-galaxy lensing.

The applicability of this work to training-set methods depends on the
details of the method.  An advantage of training set methods is that
they may ``learn'' the correct noise model automatically, and
therefore should not require any modification to reach optimum
performance (which is presumably still much reduced compared to the
no-tails case).  But for this to happen, the training set must be
sufficiently large to encompass the non-Gaussian features of the
photometry.  This may require a rather larger training set than would
otherwise be required, and it also requires a training set that is not
cleaner than the full dataset.  However, it may be possible to build a
hybrid approach in which detailed knowledge of photometry error
distributions from large sets of Monte Carlos is combined with a
modest spectroscopic sample to train the algorithm.

Non-Gaussian photometry errors may not be a substantial source of
catastrophic outliers in current surveys.  The SIM3/SIM4 tails may be
unrealistically heavy, as there is scant published data on the size of
the non-Gaussian tails for faint galaxy photometry.  Furthermore,
catastrophic outliers exist even with purely Gaussian photometry
errors, due to color-space degeneracies.  However, real-world
experience such as that of Cameron \& Driver (2007) and, in a
different context, Bolton \etal\ (2004), suggests that non-Gaussian
errors are often not negligible.  Color-space degeneracies are usually
{\it near}-degeneracies, and galaxies become much more likely to
scatter across a near-degeneracy if the the photometry has
non-Gaussian tails.

Our example started from an unrealistically good baseline of $S/N =
20$ in each of seven filters and $\delta z_{rms}=0.026$, so the effect
of the tails was particularly dramatic.  Surveys starting from a more
realistic performance baseline will not see such a large fractional
increase in scatter, but may still see the effect of tails in the
overall error budget.  Limiting the tails of the photometry error
distribution and using an accurate error model will reduce photometric
redshift scatter and greatly reduce the size of the spectroscopic
sample required to calibrate the scatter.

\end{document}